# Performance Analysis of Non-ideal Wireless PBFT Networks with mmWave and Terahertz Signals

Haoxiang Luo[1], Xiangyue Yang[1, 2], Hongfang Yu[1, 3], Gang Sun[1], Shizhong Xu[1], Long Luo[1]
[1] University of Electronic Science and Technology of China, Chengdu, China
[2] James Watt School of Engineering, University of Glasgow, Glasgow, UK
[3] Peng Cheng Laboratory, Shenzhen, China
Email: lhx991115@163.com, 2019190502004@std.uestc.edu.cn, {yuhf, gangsun, xsz, llong}@uestc.edu.cn

*Abstract*—Due to advantages in security and privacy, blockchain is considered a key enabling technology to support 6G communications. Practical Byzantine Fault Tolerance (PBFT) is seen as the most applicable consensus mechanism in blockchain-enabled wireless networks. However, previous studies on PBFT do not consider the channel performance of the physical layer, such as path loss and channel fading, resulting in research results that are far from real networks. Additionally, 6G communications will widely deploy high frequency signals such as millimeter wave (mmWave) and terahertz (THz), while the performance of PBFT is still unknown when these signals are transmitted in wireless PBFT networks. Therefore, it is urgent to study the performance of non-ideal wireless PBFT networks with mmWave and THz signals, so as to better make PBFT play a role in 6G era. In this paper, we study and compare the performance of mmWave and THz signals in non-ideal wireless PBFT networks, considering Rayleigh Fading (RF) and close-in Free Space (FS) reference distance path loss. Performance is evaluated by *consensus success rate* and *delay*. Meanwhile, we find and derive that there is a maximum distance between two nodes that can make PBFT consensus inevitably successful, and it is named *active distance* of PBFT in this paper. The research results not only analyze the performance of non-ideal wireless PBFT networks, but also provide an important reference for the future transmission of mmWave and THz signals in PBFT networks.

*Index Terms*—Blockchain, PBFT, terahertz signals, mmWave signals, 6G communications

## I. Introduction

Since the perfect combination of cryptography and consensus, blockchain is considered as a revolutionary distributed system, which provides users with a decentralized architecture and strong tamper-proof capability. Blockchain is believed to have the potential to transform the way we share information and reshape society in the future. In the information and communication area, it is also expected to protect wireless networks security in 6G communications [1-2]. Recently, it has been widely used in the Internet of Things (IoT) [3], Internet of Medical Things (IoMT) [4], Internet of Vehicles (IoV) [5], Internet of Drones (IoD) [6], and other network fields.

Therefore, it can be said that the emergence of blockchain paves the way for wireless networks in the future. The consensus mechanism in the blockchain is the basis that allows nodes in the network to establish trust without the involvement of any trusted third party. The consensus mechanism widely used in the consortium blockchain is PBFT [7], which is based on voting. PBFT has high throughput and low computational requirements, and provides 1/3 fault tolerance of the wireless network, that is, at most ($n$-1)/3 Byzantine nodes ($n$ is the total number of nodes) are allowed in the system. These features make it attractive for future wireless networks.

However, the current research on the performance of PBFT wireless networks is quite ideal. In [5], authors analyze the relationship between PBFT consensus success rate and delay in the ideal channel of IoV. In [8], authors analyze the performance of wireless PBFT networks using IEEE 802.11 protocol. [9] studies the security of PBFT in the case of sharding. And in [10], the authors study how to minimize the number of replicas to ensure PBFT consensus liveness. Despite these researches, there are still a number of issues need to be addressed to build a wireless network that supports blockchain. An important challenge is that wireless channels in the physical layer often suffer various channel fading and path losses. Channel fading can increase the bit error rate of signals, and path loss can affect the received power of the receiving node, thus, they enhance the uncertainty of wireless connections and affect the overall performance of blockchain. As a result, we need study and analyze the performance of non-ideal wireless PBFT networks. The path loss and channel fading models of wireless networks are not only related to the environment, but also to the signal frequency in the channel. In 6G communications, mmWave (26.5-100GHz) and THz (0.1-10THz) signals are regarded as important potential schemes because they provide more spectrum resources. However, since their high frequencies, there are some problems such as large propagation attenuation and short transmission distance [1]. Their performances in various scenarios are worthy of further study. Therefore, the motivation of this paper is to investigate the *consensus success rate*, and *delay* considering mmWave and THz signals transmitted in non-ideal wireless PBFT networks.

In this paper, the channel fading and path loss models we consider are RF and FS, respectively. Under the influence of them, we study the various processes of PBFT consensus, specifically, *pre-prepare*, *prepare*, *commit* and *reply*. Our work

This research was supported in part by the Natural Science Foundation of Sichuan Province (2022NSFSC0913); and in part by the PCL Future Greater-Bay Area Network Facilities for Large-scale Experiments and Applications (2018KP001).
Hongfang Yu is the corresponding author.

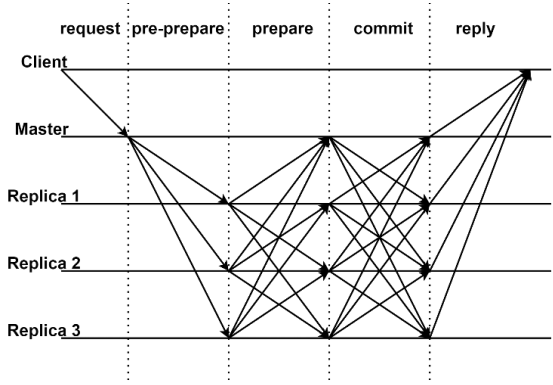

Fig. 1. PBFT consensus process.

thus demonstrates the performance of various processes for non-ideal wireless PBFT networks with mmWave and THz signals, and the factors that may affect their performance. To our best knowledge, this is the first work on performance analysis for mmWave and THz signals in PBFT networks.

The contribution of this paper can be summarized as follows
- First, we comb and analyze the Signal to Noise Ratio (SNR) and transmission success rate of mmWave and THz signals in RF and FS models.
- Second, we derive the performance of each stage of non-ideal wireless PBFT networks with mmWave and THz signals, and the performance difference between these two signals is compared by comprehensive simulations.
- Third, we find a maximum distance between two nodes, named *active distance* of PBFT. If the distance between any two nodes is less than this active value, the PBFT consensus will inevitably succeed.

The remaining contents of this paper are arranged as follows. Section II is the system model, which introduces the fundamentals of the PBFT consensus, as well as the RF and FS models. In Section III, the *consensus success rate*, and *delay* of non-ideal wireless PBFT consensus networks with mmWave and THz signals are analyzed and mathematically derived. Then, we simulate the numerical results of these above performances in Section IV. Finally, Section V is conclusion of this work.

## II. SYSTEM MODEL

In this section, we introduce the PBFT consensus, as well as the RF and FS models with THz and mmWave signals in turn.

### A. PBFT Consensus

Assuming that the wireless PBFT networks consist of $n$ nodes, for a successful consensus, there should be no more than $f$ Byzantine nodes, where $f$ is related to $n$ as follows.

$$f \leq \left\lfloor \frac{n-1}{3} \right\rfloor \quad (1)$$

As long as $f$ and $n$ meet (1), the safety and liveness of PBFT can be satisfied. However, according to the setting of PBFT, when the total number of nodes is greater than $3f+1$, the performance of PBFT networks will not be improved, but the consensus efficiency will be reduced. Therefore, we assume that the number of nodes in the wireless PBFT networks satisfies $n=3f+1$.

Before the PBFT starts consensus, it selects the primary node through a process called view configuration, and the other nodes serve as replicas. After the primary node selection, the client sends a *request* to the primary node to enter PBFT consensus process. In the case of a functioning PBFT networks, a complete consensus process is divided into four stages: *pre-prepare*, *prepare*, *commit*, and *reply* (as shown in Fig. 1). Each node in the wireless PBFT networks should participate in the following consensus process.

**Pre-prepare:** The primary node broadcasts *pre-prepare* message to all replicas.

**Prepare:** Each replica that receives the *pre-prepare* message broadcasts the *prepare* message to other replicas. If the replica receives *2f* or more *prepare* messages corresponding to the *pre-prepare* message, this *prepare* message is considered valid.

**Commit:** If the replica determines that the *prepare* message is true, it will broadcast the *commit* message to other replicas.

**Reply:** Each replica returns a *reply* message to the client as the result of the reply message.

It is important to note that the result of the request is valid only if the client receives at least $f+1$ same *replies*.

### B. RF and FS Models with THz and mmWave Signals

The characteristics of wireless communication channels determine the upper limit of the wireless communication system's performance. In order to make mmWave and THz signals better serve 6G communications, it is necessary to study the channel fading and path loss model.

First, we assume that nodes obey a two-dimensional Poisson distribution with density $\gamma$. Then, we randomly select a node as the sending node, taking it as the circle's center, and receiving nodes are distributed in the area with radius $R$. According to the two-dimensional Poisson distribution, the probability density function of the distance $r$ between the sending node and the receiving node can be expressed as

$$f(r) = \frac{d(r^2/R^2)}{dr} = \frac{2r}{R^2} \quad (2)$$

When the channels in wireless PBFT networks conform to RF, these channels are Rayleigh channels. Based on the characteristics of RF fading in wireless communications, the SNR at the receiving node is

$$SNR = \frac{P_T h r^{-\alpha}}{P_N} \quad (3)$$

where, $P_T$ is node's transmit power; $h$ represents a non-negative random variable of power gain in RF, which follows a negative exponential distribution with exponent 1. $\alpha$ represents the path loss exponent; $P_N$ is the interference noise power. We find that the parameters in (4) are all constant except $\alpha$. $\alpha$ is a variable related to the path loss. Therefore, in the next step, we need to analyze the path loss model for mmWave and THz signals.

Second, we assume that the path loss model for mmWave and THz signals is FS. According to [11], the path loss on a specific distance can be expressed as the logarithmic distance

$$PL(r)_{av} = PL(r_0) + 10\alpha \log\left(\frac{r}{r_0}\right) + X_\sigma \quad (4)$$

where, $PL(r)_{av}$ is the average path loss at distance $r$; $PL(r_0)$ represents the path loss at reference distance $r_0$ according to FS model; $X_\sigma$ is a zero-mean Gaussian random variable with standard deviation $\sigma$. In this paper, we adopt path-loss exponent ($\alpha$=2.229) with 0.22THz signals in [11], and path-loss exponent ($\alpha$=1.7) with 28GHz signals in [12], respectively.

## III. NON-IDEAL WIRELESS PBFT NETWORKS WITH THZ AND MMWAVE SIGNALS

In this section, we analyze the *consensus success rate*, and *delay* of wireless PBFT networks successively.

### A. Consensus Success Rate

We set the SNR threshold at which nodes can recover signals as $z$, then according to the two-dimensional Poisson distribution [13], the average success probability of node transmission is

$$P_s = \int_0^R P\{SNR > z\} f(r) dr = \frac{2\pi\gamma}{n} \int_0^{\sqrt{n/(\pi\gamma)}} \exp\left\{\frac{-P_N r^\alpha z}{P_T}\right\} r dr \quad (5)$$

After the PBFT completes the view configuration, the communication will be divided into four steps: *pre-prepare*, *prepare*, *commit* and *reply*, thus, we analyze the success rate of these four steps in turn.

*1) Success rate of pre-prepare:* At this stage, after receiving *request* from a client, the primary node broadcasts *pre-pare* to every replicas. According the fault tolrance of PBFT, this stage allows a maximum of $f$ communication failures. Therefore, the success rate of *pre-prepare* is

$$P_{pre-prepare} = \sum_{i=0}^{f} C_{n-1}^i (1-P_s)^i P_s^{(n-1-i)} \quad (6)$$

*2) Success rate of prepare:* Given the success rate (6) at the *pre-prepare* stage, $n$-1-$i$ nodes receive the *pre-prepare* message from the primary node. Then, each replica broadcasts *prepare* message to other replicas. To ensure successful completion of this stage, a maximum of $f$-$i$ communication failures are allowed, because *pre-prepare* stage has $i$ failures. And the success rate of *prepare* is

$$P_{prepare} = \sum_{j=0}^{f-i} C_{n-1-i}^j (1-P_s)^j P_s^{(n-1-i-j)} \quad (7)$$

*3) Success rate of commit:* This stage is very similar to prepare. The only difference is that the primary noed also need broadcast *commit* message. After the first two phases, $i+j$ nodes have not received messages properly. Therefore, the success rate of *commit* is

$$P_{commit} = \sum_{k=0}^{f-i-j} C_{n-i-j}^k (1-P_s)^k P_s^{(n-i-j-k)} \quad (8)$$

*4) Success rate of reply:* To ensure that the *reply* stage is valid, the client must receive $2f+1$ *reply* messages. In other words, this stage allows a maximum of $f$-$i$-$j$-$k$ communication failures, thus the success rate of *reply* stage is

$$P_{reply} = \sum_{l=0}^{f-i-j-k} C_{n-i-j-k}^l (1-P_s)^l P_s^{(n-i-j-k-l)} \quad (9)$$

In general, the consensus success rate of wireless PBFT networks is closely related to (6), (7), (8), and (9) and can be expressed as (10).

### B. Delay

Based on the above analysis of PBFT consensus, its communication is divided into four stages, thus, the delay of wireless PBFT networks is the sum of the four-stage communication delays. In each stage, the relationship between the communication delay and $P_s$ can be known from [5], and [14], namely

$$1-P_s = f_Q\left(\frac{NTBC - NTBR + \frac{\log NTB}{2}}{(\log e)\sqrt{NTB}}\right) \quad (11)$$

where, $f_Q$ is the $Q$ function, and $T$ represents delay for a channel. $N$ represents the number of subcarriers, in our paper $N = 1$. $B$ is the bandwidth. $R$ and $C$ are the transmission speed and channel capacity, respectively.

For the *pre-prepare* stage, the primary node need send messages to $n$-1 replicas, thus, the communication delay $t_{pre-prepare}$ can be expressed as

$$t_{pre-prepare} = (n-1)T \quad (12)$$

Similarly, in the *prepare* and *commit* stages, each node needs to broadcast messages to $n$-1 nodes, thus, $t_{prepare}$ and $t_{commit}$ are consistent with $t_{pre-prepare}$, which can be represented by $t_1$.

$$t_1 = t_{pre-prepare} = t_{prepare} = t_{commit} = (n-1)T \quad (13)$$

In addition, in the *reply* stage, each node only needs to send a *reply* message to the primary node, thus, its delay $t_{reply}$ is equal to $T$, which can be represented by $t_2$.

$$t_2 = t_{reply} = T \quad (14)$$

According to (11), we can calculate $t_1$ and $t_2$, respectively. Furthermore, we can obtain that the total delay of wireless PBFT networks is

$$t_{total} = t_3 = 3t_1 + t_2 \quad (15)$$

## IV. NUMERICAL SIMULATION

Before numerical simulations, we need to set the values of parameters related to the above performances in wireless PBFT networks, which are shown in Table I. In addition, in order to study the effects of the SNR threshold $z$, and node density $\gamma$ on the performances, we assume three sets of data to get a more comprehensive reference, which are $z$=6 dB, $\gamma$=2 nodes/m²; $z$=6 dB, $\gamma$=5 nodes/m²; $z$=4 dB, $\gamma$=5 nodes/m².

### A. Simulation of Success Rate

First, we simulate the transmission success rate of mmWave and THz signals in RF and FS models. The simulation results are shown in Fig. 2 (a). As the number of nodes increases, $P_s$ will decrease. The reason is that, according to the two-dimensional Poisson distribution, when there are too many nodes in wireless networks, the distance between some nodes become too large. Then, with the increase of distance, the influence of RF and FS is more obvious. Additionally, the

$$P_c = \sum_{i=0}^{f}\left(C_{n-1}^i (1-P_s)^i P_s^{(n-1-i)} \sum_{j=0}^{f-i}\left(C_{n-1-i}^j (1-P_s)^j P_s^{(n-1-i-j)} \sum_{k=0}^{f-i-j}\left(C_{n-i-j}^k (1-P_s)^k P_s^{(n-i-j-k)} \sum_{l=0}^{f-i-j-k} C_{n-i-j-k}^l (1-P_s)^l P_s^{(n-i-j-k-l)}\right)\right)\right) \quad (10)$$

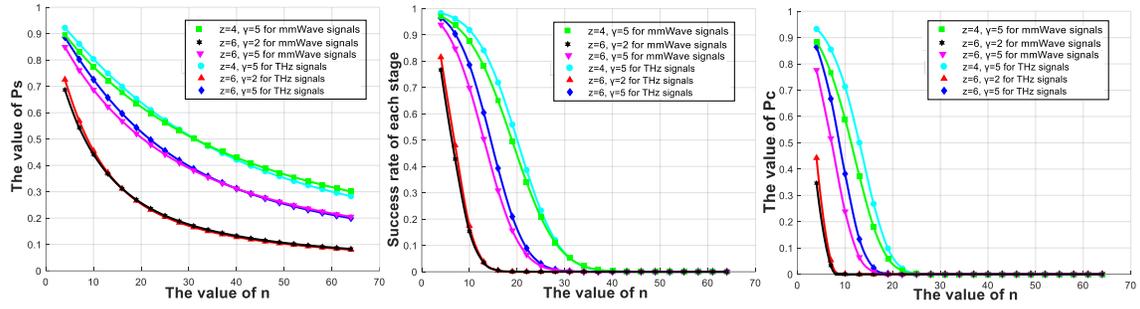

Fig. 2. (a) The value of $P_s$; (b) Success rate of each stage; (c) The value of $P_c$.

values of $z$ and $\gamma$ also affect the $P_s$ values. A lower value of $z$ indicates that the receiving node has a stronger capability to recover signal, which contributes to the transmission success rate. The lower the value of $\gamma$ indicates that the distance between nodes increases, leading to a decrease in the transmission success rate. Moreover, under the same $z$ and $\gamma$ values, when the number of nodes is small, the performance of mmWave signals is worse than THz signals. And when the number of nodes increases, the performance of mmWave signals is better than THz signals. The reason for this is that THz is a higher frequency signal than mmWave, and is more likely to be negatively affected by distance.

Second, we simulate the success rate of each stage in wireless PBFT networks. The results are shown in Fig. 2 (b). When the number of nodes is small, the fault tolerance of PBFT can improve the transmission success rate compared with Fig. 2 (a). However, when the $n$ value is large, the transmission success rate significantly decreases, indicating that the increase of distance between nodes has a negative impact on the communication ability of wireless PBFT networks. In addition, under the same $z$ and $\gamma$ values, the success rate of THz signals is higher than that of mmWave signals.

Third, we simulate the consensus success rate $P_c$ of wireless PBFT networks, as shown in Fig. 2 (c). This result is similar to Fig. 2 (b), except that the value of $P_c$ is decreased compared with Fig. 2 (b). It is quite understandable that $P_c$ is the probability of simultaneous success in all four stages of PBFT. Additionally, the decrease of $P_c$ value with the increase of $n$ indicates that mmWave and THz signals are not suitable for communication in long-distance wireless PBFT networks. This result just shows that high-frequency signals such as mmWave and THz are affected by spatial distance easily.

After the above simulation about success rate, we find that on the one hand, the SNR threshold $z$ of the receiving node can be reduced to improve the success rate; On the other hand, it can improve the node density to increase the success rate. Moreover, we also find that the distance between nodes is an important factor affecting the transmission success rate of wireless PBFT network, thus, we hope to explore a maximum distance to make PBFT consensus inevitably successful. As a result, further to ensure that the receiving node can recover the signal, its SNR threshold $z$ should be less than or equal to the SNR of signal, namely

TABLE I.    PARAMETER VALUES

| Parameters | | Values |
|---|---|---|
| THz signal | $P_N$ | 0.2 W |
| | $P_T$ | 1 W |
| | $B$ | 10 GHz |
| | $C$ | 80 Gbps |
| | $R$ | 40 Gbps |
| | $\alpha$ | 2.229 [11] |
| mmWave signal | $P_N$ | 0.2 W |
| | $P_T$ | 1 W |
| | $B$ | 800 MHz |
| | $C$ | 8 Gbps |
| | $R$ | 4 Gbps |
| | $\alpha$ | 1.7 [12] |

$$z \leq SNR = \frac{P_T h r^{-\alpha}}{P_N} \quad (16)$$

Then, we can obtain the relationship between $r$ and other parameters in (17).

$$r \leq \left(\frac{P_T h}{z P_N}\right)^{-\alpha} \quad (17)$$

If the distance between any two nodes in wireless PBFT networks satisfies (17), then the PBFT consensus must be successful. And this distance is named the *active distance*.

### B. Simualtion of Delay

For **THz signals**, Fig. 3 (a) shows the relationship between $t_1$ and the number of nodes $n$ with different values of $z$ and $\gamma$. $t_1$ shows a linear growth trend with the increase of $n$, and the values of $z$ and $\gamma$ have little influence on the change of $t_1$. This indicates that broadcasting messages to multiple nodes at the same time can prolong the delay. Fig. 3 (b) shows the relationship between $t_2$ and $n$ with different values of $z$ and $\gamma$. $t_2$ does not show a regular change with the increase of $n$, but shows a fluctuation characteristic. Meanwhile, with different values of $z$ and $\gamma$, the fluctuation law of $t_2$ is not completely consistent. And the fluctuation range always stays in 0.038-0.04as (1as=$10^{-18}$s). This result indicates that $t_2$ is one or two orders of magnitude smaller than $t_1$, indicating that $t_1$ plays a decisive role in the total delay. As a result, the delay of the first three stages (*pre-prepare*, *prepare*, *commit*) accounts for most of the total delay, while the delay of the last stage (*reply*) is so short that it can be ignored. In Fig. 3 (c), $t_{total}$ shows a change trend similar to $t_1$. It grows linearly with the increase of $n$. Its value is in the order of as, which shows the high-speed character of THz signal. Moreover, when THz signals are transmitted fast in wireless PBFT networks the simulation results also show that the number of nodes plays an important role in the delay.

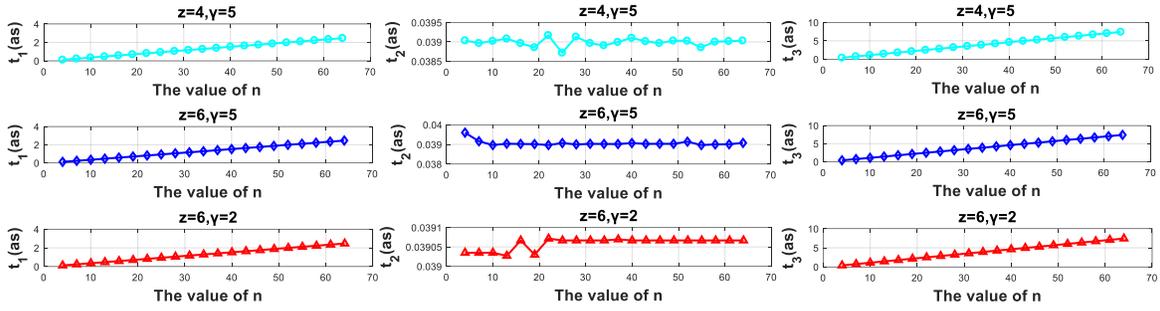

Fig. 3. (a) The value of $t_1$ with THz signals; (b) The value of $t_2$ with THz signals; (c) The value of $t_{total}$ with THz signals.

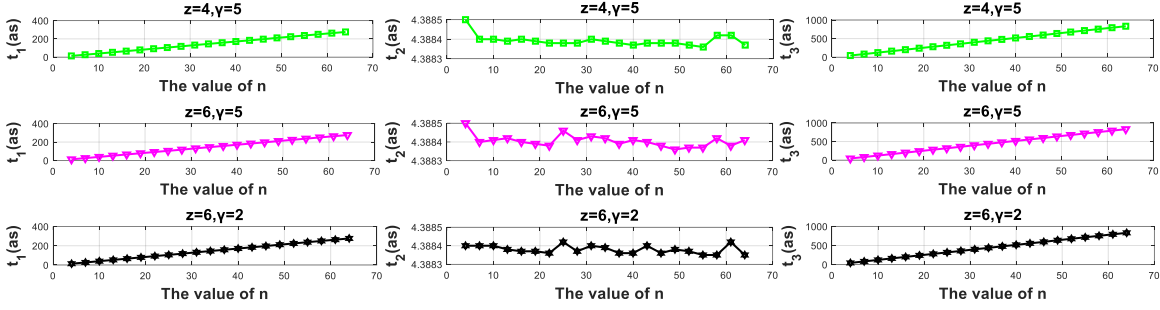

Fig. 4. (a) The value of $t_1$ with mmWave signals; (b) The value of $t_2$ with mmWave signals; (c) The value of $t_{total}$ with mmWave signals.

For **mmWave signals**, Fig. 4 (a), (b), and (c) show the characteristics of $t_1$, $t_2$, and $t_{total}$. $t_1$ and $t_{total}$ have similar properties to THz signals, but two orders of magnitude more than THz signals. This indicates that THz can provide more bandwidth than mmWave, thus, it has higher communication rates. And $t_2$ also shows an irregular fluctuation characteristic, and the fluctuation range is from 4.3883-4.3885as.

## V. CONCLUSION

In this paper, we analyze the performance of non-ideal wireless PBFT networks with mmWave and THz signals, including consensus success rate and delay. We deduce the theoretical calculation methods of the above performances in turn, and further carry out numerical simulations. Simulation results show that THz is superior to mmWave in terms of delay, because THz can provide more spectrum resources. In addition, when the node density is not large enough, THz also has a higher consensus success rate than mmWave. Moreover, we derive the maximum distance between any two nodes, called the *active distance*, which can make PBFT inevitably successful. The above results can provide a valuable reference for the practical deployment of PBFT in 6G communications.

## REFERENCES

[1] V. -L. Nguyen, P. -C. Lin, B. -C. Cheng, *et al.*, "Security and Privacy for 6G: A Survey on Prospective Technologies and Challenges," in *IEEE Communications Surveys & Tutorials*, vol. 23, no. 4, pp. 2384-2428, 2021.

[2] B. Cao, Z. Zhang, D. Feng, *et al.*, "Performance analysis and comparison of PoW, PoS and DAG based blockchains," in *Digital Communications and Networks*, vol. 6, no. 4, pp. 480-485, 2020.

[3] X. Xu, G. Sun and H. Yu, "An Efficient Blockchain PBFT Consensus Protocol in Energy Constrained IoT Applications," *2021 International Conference on UK-China Emerging Technologies (UCET)*, 2021, pp. 152-157.

[4] Y. Chen, H. Luo and Q. Bian, "A Privacy Protection Method Based on Key Encapsulation Mechanism in Medical Blockchain," *2021 IEEE 21st International Conference on Communication Technology (ICCT)*, 2021, pp. 295-300.

[5] X. Yang, H. Luo, J. Duan and H. Yu, "Ultra Reliable and Low Latency Authentication Scheme for Internet of Vehicles Based on Blockchain," *IEEE INFOCOM 2022 - IEEE Conference on Computer Communications Workshops (INFOCOM WKSHPS)*, 2022, pp. 1-5.

[6] H. Luo, S. Liu, S. Xu, J. Luo, " LECast: A Low-Energy-Consumption Broadcast Protocol for UAV Blockchain Networks, " in *Drones*, vol. 7, pp. 1-20, 2023.

[7] M. Castro, B. Liskov, "Practical byzantine fault tolerance," *OsDI* 1999, vol. 99, pp. 173-186.

[8] Z. Zhou, O. Onireti, L. Zhang and M. A. Imran, "Performance Analysis of Wireless Practical Byzantine Fault Tolerance Networks Using IEEE 802.11," *2021 IEEE Globecom Workshops (GC WKSHPS)*, 2021, pp. 1-6.

[9] X. Li, H. Luo, J. Duan, "Security Analysis of Sharding in Blockchain with PBFT Consensus," *2022 4th International Conference on Blockchain Technology*, 2022, pp. 9-14.

[10] O. Onireti, L. Zhang and M. A. Imran, "On the Viable Area of Wireless Practical Byzantine Fault Tolerance (PBFT) Blockchain Networks," *2019 IEEE Global Communications Conference (GLOBECOM)*, 2019, pp. 1-6.

[11] N. A. Abbasi, A. Hariharan, A. M. Nair and A. F. Molisch, "Channel Measurements and Path loss Modeling for Indoor THz Communication," *2020 14th European Conference on Antennas and Propagation (EuCAP)*, 2020, pp. 1-5.

[12] 3GPP TR 38.901 v.15.0.0, "Study on channel model for frequencies from 0.5 to 100 GHz," 2018.

[13] J. G. Andrews, F. Baccelli and R. K. Ganti, "A Tractable Approach to Coverage and Rate in Cellular Networks," in *IEEE Transactions on Communications*, vol. 59, no. 11, pp. 3122-3134, 2011.

[14] B. Chang, L. Zhang, L. Li, G. Zhao and Z. Chen, "Optimizing Resource Allocation in URLLC for Real-Time Wireless Control Systems," in *IEEE Transactions on Vehicular Technology*, vol. 68, no. 9, pp. 8916-8927, 2019.